\documentclass[prd,aps,groupedaddress,showpacs,nobibnotes]{revtex4}
\usepackage{epsf,epsfig,graphicx}

\newcounter{muni}

\begin{document}
\hbadness=10000 \pagenumbering{arabic}

\title{Resummation with Wilson lines off the light cone}

\author{Hsiang-nan Li$^{1}$}
\email{hnli@phys.sinica.edu.tw}

\affiliation{$^{1}$Institute of Physics, Academia Sinica, Taipei,
Taiwan 115, Republic of China,}

\affiliation{$^{1}$Department of Physics, Tsing-Hua University,
Hsinchu, Taiwan 300, Republic of China,}

\affiliation{$^{1}$Department of Physics, National Cheng-Kung
University, Tainan, Taiwan 701, Republic of China}

\begin{abstract}

I review the resummation formalism for organizing large logarithms in
perturbative expansion of collinear subprocesses through the variation
of Wilson lines off the light cone. A master equation is derived, which
involves the evolution kernel resulting from this variation. It is
then demonstrated that all the known single- and double-logarithm
summations for a parton distribution function or a transverse-momentum-dependent
parton distribution can be reproduced from the master equation by
applying appropriate soft-gluon approximations to the evolution kernel.
Moreover, jet substructures, information which is crucial for particle
identification at the Large Hadron Collider and usually acquired from
event generators, can also be calculated in this formalism.

\end{abstract}

\maketitle

\section{INTRODUCTION}

It is known that radiative corrections in perturbative QCD (pQCD) produce
large logarithms at each order of the coupling constant. Double logarithms
appear in processes involving two scales, such as $\ln^2(p^+b)$ with
$p^+$ being the large longitudinal momentum of a parton and $b$ being the
impact parameter conjugate to the small parton
transverse momentum $k_T$. In the region with a large Bjorken
variable $x$, there exists $\ln^2(1/N)$ from the Mellin transformation of
$\ln(1-x)/(1-x)_+$, for which the two scales are the large $p^+$ and the
small infrared cutoff $(1-x)p^+$ for gluon emissions from a parton.
Single logarithms are generated in processes involving one scale, such
as $\ln p^+$ and $\ln(1/x)$, for which the relevant scales are the large
$p^+$ and the small $xp^+$, respectively. To improve
perturbative expansion, these logarithmic corrections need to be organized by
evolution equations or resummation techniques. Various methods have been
developed to organize these logarithmic corrections to a parton distribution
function (PDF) or to a transverse-momentum-dependent distribution function (TMD):
the $k_T$ resummation for $\ln^2(p^+b)$ \cite{CS,Collins:1984kg}, the
threshold resummation for $\ln^2(1/N)$ \cite{S,CT,KM}, the joint resummation
\cite{Li99,LSV00} that unifies the above two formalisms, the
Dokshitzer-Gribov-Lipatov-Altarelli-Parisi (DGLAP)  equation
for $\ln p^+$ \cite{AP}, the Balitsky-Fadin-Kuraev-Lipatov (BFKL)
equation for $\ln(1/x)$ \cite{BFKL}, and
the Ciafaloni-Catani-Fiorani-Marchesini (CCFM) equation \cite{CCFM}
that combines the above two evolution equations.

The definition of a PDF or a TMD contains Wilson lines along the light
cone, which collect gluons collimated to a beam particle of momentum $p$
and attaching to other parts of a scattering process. The Wilson lines contain
vertical links at infinity, if a TMD is considered. To perform resummation,
a simple trick is to vary the Wilson lines off the light cone into an
arbitrary direction $n^\mu$ with $n^2\not=0$ \cite{Li:1995eh}. The
PDF or the TMD must depend on $p^\mu$ and $n^\mu$ through the Lorentz
invariants $p\cdot n$ and $n^2$. When a parton $k_T$ is involved,
the phase space of real radiation is constrained, so the associated infrared
enhancement does not cancel completely that in virtual correction. The
infrared enhancement then generates the double logarithms of the ratio
$(p\cdot n)^2/(k_T^2n^2)$, and the variation of $n$ turns into the variation
of the scale $p^+$ or $k_T$. The key is that all different choices of the
vector $n$ are equivalent in the viewpoint of collecting the collinear divergences
associated with the beam particle. Therefore, the effect from varying $n$
does not involve the collinear divergences, which can then be factorized
out of the PDF or the TMD, leading to an evolution equation in $n$. The
resummation technique via the variation of the Wilson lines off the light
cone will be reviewed in this article, and its wide applications to the
single- and double-logarithmic summations will be demonstrated.

It has been a long-standing challenge to predict substructures
(including masses and energy profiles) of light-quark and gluon jets
in the pQCD theory: fixed-order QCD calculations
cannot describe experimental data on jet substructures, especially in
extreme kinematic regions, such as the region with a small jet invariant
mass. Hence, it is a custom for experimentalists to compare measured jet
substructures with predictions from full event generators such as PYTHIA or
HERWIG. While the full event generators (usually with specific tuning)
could describe data, it remains desirable to develop a theoretical
framework for the study of jet substructures. A novel approach to
predicting jet substructures based on the resummation formalism
was proposed recently \cite{Li:2011hy}. It has been shown that results of this
formalism for light-quark and gluon jets are well consistent with the mass
distributions measured by CDF \cite{Aaltonen:2011pg}, and with the
energy profiles measured by CDF at Tevatron  \cite{Acosta:2005ix} and CMS
at Large Hadron Collider (LHC) \cite{CMSJE}.

\section{Resummation Formalism}

In this section I explain the basic idea of the resummation formalism with
the Wilson lines off the light cone. Collinear and soft divergences in
perturbation theory may overlap to form double logarithms in extreme
kinematic regions with low $p_T$ and large $x$. The former includes low $p_T$
jet, photon, and $W$ boson productions, which all require real gluon emissions
of small $p_T$. The latter includes top pair production, deeply inelastic
scattering (DIS), Drell-Yan production,
and heavy meson decays $B\to X_u l\nu$ and $B\to X_s\gamma$ \cite{LY1,KS,L1}
at the end points, for which parton momenta remain large, and radiations
are constrained in the soft region. Because of the limited phase space
for real corrections, the infrared cancellation is not complete. The double
logarithms, appearing in products with the coupling constant $\alpha_s$, such as
$\alpha_s\ln^2(E/p_T)$ with the beam energy $E$ and $\alpha_s\ln(1-x)/(1-x)_+$,
deteriorate perturbative expansion. Double logarithms also occur in exclusive
processes, such as Landshoff scattering \cite{BS}, hadron form factors
\cite{LS}, Compton scattering \cite{CL93} and heavy-to-light transitions
$B\to\pi(\rho)$ \cite{LY2} and $B\to D^{(*)}$ \cite{L2} at maximal recoil.
In order to have a reliable pQCD analysis of these processes,
the important logarithms need to be summed to all orders.

\begin{figure}
\begin{center}
\centering\includegraphics[width=0.6\linewidth]{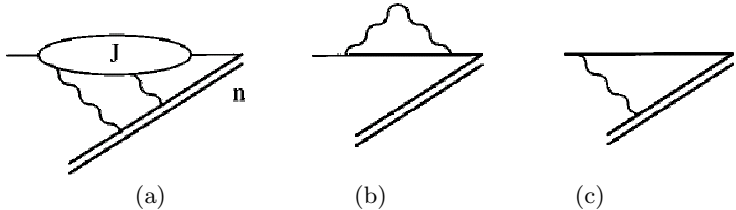}

(a)\hspace{2.5cm}(b)\hspace{2.5cm}(c)
\end{center}
\caption{(a) Jet subprocess defined in Eq.~(\ref{j}).
(b) and (c) LO diagrams of (a). \label{resum}}
\end{figure}

Take as an example a jet subprocess defined by the matrix element in the covariant
gauge $\partial\cdot A=0$ \cite{L1},
\begin{equation}
J(p,n)u(p)=\langle 0|{\cal P}\exp\left[-ig\int_0^\infty dz n\cdot A(nz)\right]
q(0)|p\rangle\;,
\label{j}
\end{equation}
where $q$ is a light quark field with momentum $p$, $u(p)$ is a spinor,
and $A$ is a gluon field.
The abelian case of this subprocess has been discussed in \cite{C}. The
path-ordered exponential in Eq.~(\ref{j}) is the consequence of the
factorization of collinear gluons with momenta parallel to $p$
from a full process. For convenience, it is assumed that $p$ has a large
light-cone component $p^+$, and all its other components vanish. A general
diagram of the jet function $J$ is shown in Fig.~\ref{resum}(a), where the
path-ordered exponential is represented by a double line along the arbitrary
vector $n$. As stated before, varying the direction $n$ does not change the
collinear divergences collected by the Wilson line.

It is easy to see that $J$ contains double logarithms from the overlap of
collinear and soft divergences by calculating the leading-order (LO) diagrams in
Fig.~\ref{resum}(b), the self-energy correction, and in Fig.~\ref{resum}(c),
the vertex correction. In the covariant gauge Fig.~\ref{resum}(b)
(Fig.~\ref{resum}(c)) produces a single (double) logarithm. In the axial gauge $n\cdot A=0$
the path-ordered exponential reduces to an identity, and Fig.~\ref{resum}(c)
does not exist. The essential step in the resummation technique is to derive a
differential equation $p^+dJ/dp^+=CJ$ \cite{L1}, where the
coefficient function $C$ contains only single logarithms, and can be
treated by renormalization-group (RG) methods. Since the path-ordered
exponential is scale-invariant in $n$, $J$ must depend on $p$ and $n$ through
the ratio $(p\cdot n)^2/n^2$. The differential operator $d/dp^+$
can then be replaced by $d/dn$ using a chain rule
\begin{equation}
p^+\frac{d}{dp^+}J=-\frac{n^2}{v\cdot n}v_\alpha\frac{d}{dn_\alpha}J,
\label{cr}
\end{equation}
with the vector $v=(1,0,{\bf 0}_T)$ being defined via $p=p^+v$.

\begin{figure}
\centering\includegraphics[width=0.5\linewidth]{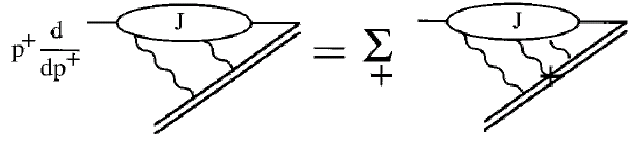}
\caption{Derivative $p^+dJ/dp^+$ in the covariant gauge. \label{resum2}}
\end{figure}

Equation (\ref{cr}) simplifies the analysis tremendously, because $n$
appears only in the Feynman rules for the Wilson line, while $p$ may flow
through the whole diagram in Fig.~\ref{resum}(a). The differentiation of each
eikonal vertex and of the associated eikonal propagator with respect to
$n_\alpha$,
\begin{eqnarray}
-\frac{n^2}{v\cdot n}v_\alpha\frac{d}{dn_\alpha}\frac{n_\mu}{n\cdot l}
=\frac{n^2}{v\cdot n}\left(\frac{v\cdot l}{n\cdot l}n_\mu-v_\mu\right)
\frac{1}{n\cdot l}\equiv\frac{{\hat n}_\mu}{n\cdot l},
\label{dp}
\end{eqnarray}
leads to the special vertex ${\hat n}_\mu$. The derivative $p^+dJ/dp^+$ is
thus expressed as a summation over different attachments of ${\hat n}_\mu$,
labeled by the symbol $+$ in Fig.~\ref{resum2}.
If the loop momentum $l$ is parallel to $p$, the factor $v\cdot l$
vanishes, and ${\hat n}_\mu$ is proportional to $v_\mu$. When this
${\hat n}_\mu$ is contracted with a vertex in $J$, in which all momenta
are mainly parallel to $p$, the contribution to $p^+dJ/dp^+$ is
suppressed. Hence, the leading regions of $l$ are soft and hard.

\begin{figure}
\centering\includegraphics[width=0.5\linewidth]{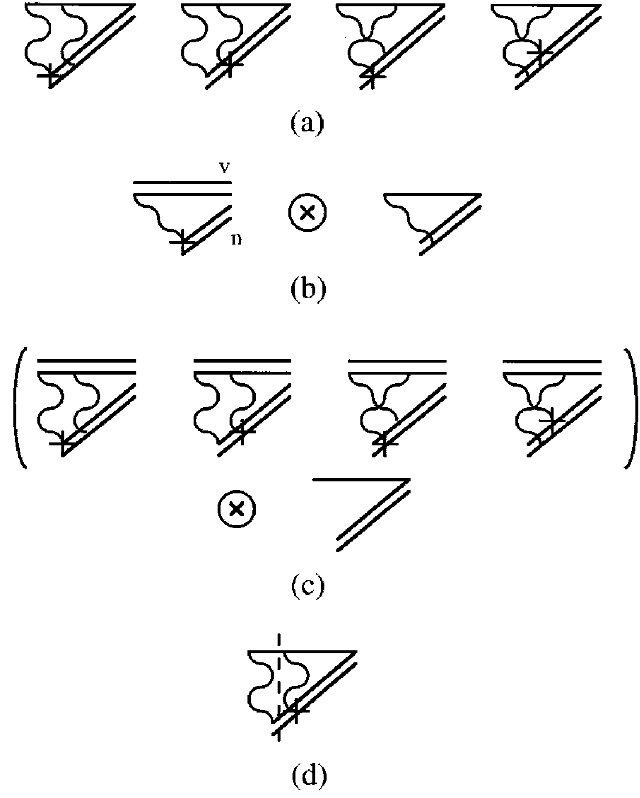}
\caption{(a) $O(\alpha_s^2)$ examples for the differentiated $J$.
(b) Factorization of $K$ at $O(\alpha_s)$.
(c) Factorization of $K$ at $O(\alpha_s^2)$.
(d) Factorization of $G$ at $O(\alpha_s)$. \label{resum3}}
\end{figure}

According to this observation, we investigate some two-loop examples
exhibited in Fig.~\ref{resum3}(a). Note that the third and fourth
diagrams in Fig.~\ref{resum3}(a), involving the crossing gluons, do not
mean three-loop diagrams. If the loop momentum flowing through the
special vertex is soft but another is not, only the first diagram is important,
giving a large single logarithm. In this soft region the subdiagram
containing the special vertex can be factorized using the eikonal approximation
as shown in Fig.~\ref{resum3}(b), where the symbol $\otimes$ represents a
convoluting relation. The subdiagram is absorbed into a soft kernel $K$, and
the remainder is identified as the original jet function $J$, both being
$O(\alpha_s)$ contributions. If both the loop momenta are soft, the four
diagrams in Fig.~\ref{resum3}(a) are equally important. The subdiagrams,
factorized according to Fig.~\ref{resum3}(c), contribute to $K$ at
$O(\alpha_s^2)$, and the remainder is the LO diagram of $J$. If the loop momentum
flowing through the special vertex is hard and another is not, the second
diagram in Fig.~\ref{resum3}(a) dominates. In this region the subdiagram
containing the special vertex is factorized as shown in Fig.~\ref{resum3}(d).
The right-hand side of the dashed line is absorbed into a hard kernel $G$
as an $O(\alpha_s)$ contribution, and the left-hand side is identified as the
$O(\alpha_s)$ diagram of $J$. If both the loop momenta are hard, all the
diagrams in Fig.~\ref{resum3}(a) are absorbed into $G$, giving the
$O(\alpha_s^2)$ contributions.

Extending the above reasoning to all orders, one derives the differential
equation
\begin{equation}
p^+\frac{d}{dp^+}J=\left[K(m/\mu,\alpha_s(\mu))+G(p^+\nu/\mu,
\alpha_s(\mu))\right]J,
\label{de1}
\end{equation}
where the coefficient function $C$ has been written as the sum of the
soft kernel $K$ and the hard kernel $G$. In the above expression $\mu$ is a
factorization scale, the gauge factor in $G$ is defined as
$\nu=\sqrt{(v\cdot n)^2/|n^2|}$, and a gluon mass $m$ has been
introduced to regularize the infrared divergence in $K$. Note that the
function $J$ defined in Eq.~(\ref{j}) is not a physical object, and is not
infrared safe, so the infrared regulator $m$ is needed \cite{C}.
For physical objects, such as the TMD discussed in the next section,
this infrared regulator is not necessary.
The $O(\alpha_s)$ contribution to $K$ from Fig.~\ref{resum3}(b) is written as
\begin{eqnarray}
K&=&-ig^2 C_F\mu^\epsilon\int\frac{d^{4-\epsilon} l}
{(2\pi)^{4-\epsilon}}\frac{{\hat n}_\mu}{n\cdot l}
\frac{g^{\mu\nu}}{l^2-m^2}\frac{v_\nu}{v\cdot l}-\delta K,
\label{k1}
\end{eqnarray}
where
\begin{eqnarray*}
\delta{K}=-\frac{\alpha_s}{2\pi}C_F\left(\frac{2}{\epsilon}
+\ln 4\pi-\gamma_E\right),
\label{dkp1}
\end{eqnarray*}
is an additive counterterm in the ${\overline {\rm MS}}$ scheme, $\gamma_E$
being the Euler constant. The $O(\alpha_s)$
contribution to $G$ from Fig.~\ref{resum3}(d) is given by
\begin{eqnarray}
G=-ig^2 C_F\mu^\epsilon\int\frac{d^{4-\epsilon} l}
{(2\pi)^{4-\epsilon}}\frac{{\hat n}_\mu}{n\cdot l}\frac{g^{\mu\nu}}{l^2}
\left(\frac{\not p+\not l}{(p+l)^2}\gamma_\nu-\frac{v_\nu}{v\cdot l}\right)
-\delta G,
\label{g1}
\end{eqnarray}
where the second term in the parentheses acts as a soft subtraction to avoid
double counting, and $\delta G$ is an additive counterterm. Because of the soft
cancellation between the two terms in the above expression, the infrared regulator
$m$ has been dropped. A straightforward
evaluation shows that Eqs.~(\ref{k1}) and (\ref{g1}) contain only the single
logarithms $\ln(m/\mu)$ and $\ln(p^+\nu/\mu)$, respectively, as expected.
Organizing these single logarithms using RG methods, and then
solving Eq.~(\ref{de1}), one resums the double logarithms $\ln^2(p^+/m)$ in $J$.

To reproduce all the known resummations and evolution equations, we
construct a master equation for the TMD $\Phi(x,k_T)$ following the above
procedures. The dependence on a factorization
scale $\mu$ is implicit. The factorization scale is similar to a
renormalization scale, but introduced in perturbative computations for
an effective theory. If the parton is a quark, $\Phi$ is defined by
\begin{eqnarray}
\Phi_{q/N}(x,k_T)&=&\int\frac{dy^-}{2\pi}\int\frac{d^2y_T}{(2\pi)^2}
e^{-ix p^+y^-+i{\bf k}_T\cdot {\bf y}_T}\nonumber\\
& &\times \frac{1}{2}\langle N(p,\sigma)|\bar q(0,y^-,{\bf y}_T)\frac{1}{2}\gamma^+
W(y^-,{\bf y}_T,0,0_T)q(0,0,0_T)|N(p,\sigma)\rangle,\label{deq}
\end{eqnarray}
where $|N(p,\sigma)\rangle$ denotes the bound state of the nucleon
with momentum $p$ and spin $\sigma$, $y=(0,y^-,{\bf y}_T)$ is the
coordinate of the quark field after the final-state cut, the
first factor $1/2$ is attributed to the average over the nucleon spin,
and the matrix $\gamma^+/2$ is the spin projector for the nucleon.
The Wilson links are defined as $W(y^-,{\bf y}_T,0,0_T)=W(0,0_T)I_{0,{\bf y}_T}
W^\dagger(y^-,{\bf y}_T)$ with the vertical link $I_{0,{\bf y}_T}$ being
located at $y^-=\infty$ \cite{BJY}, and
\begin{eqnarray}
W(y^-,{\bf y}_T)= {\cal P} \exp\left[-ig \int_0^\infty dz
n\cdot A(y+z n)\right].\label{wilson}
\end{eqnarray}
The two quark fields before and after the final-state cut in Eq.~(\ref{deq}) are
separated by a distance, so the above Wilson links are demanded by the gauge
invariance of the TMD as a nonlocal matrix element. More investigations on the
vertical Wilson links can be found in \cite{CS08}. If the parton is a gluon,
the nonlocal operator in Eq.~(\ref{deq}) is replaced by
$F^+_\mu(0,y^-,y_T)F^{\mu+}(0,0,0_T)$.

Similarly, $n$ is varied arbitrarily away from the light cone with $n^2\not= 0$.
Then $\Phi$ depends on $p^+$ via the ratio $(p\cdot n)^2/n^2$, so
the chain rule in Eq.~(\ref{cr}) relating the derivative $d\Phi/dp^+$ to
$d\Phi/dn_\alpha$ applies. One derives the master equation
\begin{eqnarray}
p^+\frac{d}{dp^+}\Phi(x,k_T)=2{\bar \Phi}(x,k_T),
\label{meq}
\end{eqnarray}
where $\bar \Phi$ contains the special vertex, and the coefficient 2 is
due to the equality of $\bar\Phi$ with the special vertex on
either side of the final-state cut.
The function $\bar \Phi$ is factorized into the convolution of the
soft and hard kernels with $\Phi$:
\begin{eqnarray}
{\bar \Phi}(x,k_T)={\bar \Phi}_{s}(x,k_T)+
{\bar \Phi}_{h}(x,k_T),
\label{ssf}
\end{eqnarray}
with the soft contribution
\begin{eqnarray}
{\bar \Phi}_{s}&=&\left[-ig^2 C_F\mu^\epsilon\int\frac{d^{4-\epsilon} l}
{(2\pi)^{4-\epsilon}}\frac{{\hat n}\cdot v}{n\cdot ll^2 v\cdot l}
-\delta K\right]\Phi(x,k_T)\nonumber\\
& &-ig^2 C_F\mu^\epsilon\int\frac{d^{4-\epsilon} l}
{(2\pi)^{4-\epsilon}}\frac{{\hat n}\cdot v}{n\cdot l v\cdot l}2\pi i\delta(l^2)
\Phi(x+l^+/p^+,|{\bf k}_T+{\bf l}_T|),
\label{fsr}
\end{eqnarray}
where the first term is the same as in Eq.~(\ref{k1}), and
the second term proportional to $\delta(l^2)$ arises from the real soft
gluon emission. The hard contribution is given by
${\bar\Phi}_h(x,k_T)=G(xp^+\nu/\mu,\alpha_s(\mu))\Phi(x,k_T)$, in which
the hard kernel $G$ is the same as in Eq.~(\ref{g1}).

\section{$k_T$ Resummation and BFKL Equation}

The TMD definition in Eq.~(\ref{deq}) contains three scales:
$(1-x)p^+$, $xp^+$, and $k_T$. We first consider the soft approximation
corresponding to the rapidity ordering of real gluon emissions in a
ladder diagram. Assume that a parton carries the longitudinal momentum
$xp^++l_2^++l_1^+$, which becomes $xp^++l_1^+$ after emitting a gluon
of longitudinal momentum $l_2^+$ and transverse momentum $l_{2T}$,
and then becomes $xp^+$ after emitting a gluon of longitudinal
momentum $l_1^+$ and transverse momentum $l_{1T}$. In the kinematic
configuration with $l_2^+\gg l_1^+$ and $l_{2T}\sim l_{1T}$,
the original parton momentum is approximated by
$xp^++l_2^++l_1^+\approx xp^++l_2^+$. The loop integral
associated with the first gluon emission is then independent of $l_1^+$,
and can be worked out straightforwardly, giving a logarithm.
The loop integral associated with the second gluon emission, involving
only $l_1^+$, also gives a logarithm. Hence, a ladder diagram with
$N$ rung gluons generates the logarithmic correction $(\alpha_s L)^N$
under the above ordering, where $L$ denotes the large
logarithm. Following the rapidity ordering, we adopt
the approximation for the real gluon emission in Eq.~(\ref{fsr})
\begin{equation}
\Phi(x+l^+/p^+,|{\bf k_T+\bf l_T}|)\approx
\Phi(x,|{\bf k_T+\bf l_T}|),
\label{nl}
\end{equation}
where the $l^+$ dependence has been neglected. The transverse momenta
$l_T$, being of the same order as $k_T$ in this kinematic configuration,
is kept. The variable $l^+$ in $K$ is then integrated up to infinity,
such that the scale $(1-x)p^+$ disappears.

Equation~(\ref{meq}) is Fourier transformed into the impact parameter $b$
space, with the definition 
$\int \Phi(x,k_T)\exp(i{\bf k}_T\cdot {\bf b})d^2k_T/(2\pi)^2\equiv \Phi(x,b)$.
The convolution in the transverse-momentum space in Eq.~(\ref{fsr}) then
becomes a product under the Fourier transformation. In the intermediate $x$ region
$\Phi$ involves two scales, the large $xp^+$ that characterizes the hard
kernel $G$ and the small $1/b$ that characterizes the
soft kernel $K$. The master equation (\ref{meq}) becomes
\begin{eqnarray}
p^+\frac{d}{dp^+}\Phi(x,b)=2\left[K(1/(b\mu),\alpha_s(\mu))+
G(xp^+\nu/\mu,\alpha_s(\mu))\right]\Phi(x,b),
\label{dph}
\end{eqnarray}
whose solution with $\nu=1$ leads to the $k_T$ resummation
\begin{eqnarray}
\Phi(x,b)=\Delta_k(x,b)\Phi_i(x),
\label{sph}
\end{eqnarray}
with the Sudakov exponential
\begin{eqnarray}
\Delta_k(x,b)=\exp\left[-2\int_{1/b}^{xp^+}\frac{d p}{p}
\int_{1/b}^{p}\frac{d\mu}{\mu}\gamma_{K}(\alpha_s(\mu))\right],
\label{fb}
\end{eqnarray}
and the initial condition $\Phi_i$ of the Sudakov evolution.
The anomalous dimension of $K$, $\lambda_K=\mu d\delta K/d\mu$,
is given, up to two loops, by \cite{KT82}
\begin{eqnarray}
\gamma_K=\frac{\alpha_s}{\pi}C_F+\left(\frac{\alpha_s}{\pi}
\right)^2C_F\left[{C}_A\left(\frac{67}{36}
-\frac{\pi^{2}}{12}\right)-\frac{5}{18}n_{f}\right]\;,
\label{lk}
\end{eqnarray}
with $n_{f}$ being the number of quark flavors and $C_A=3$ being a color factor.

\begin{figure}
\centering\includegraphics[angle=-90,width=.7\linewidth]{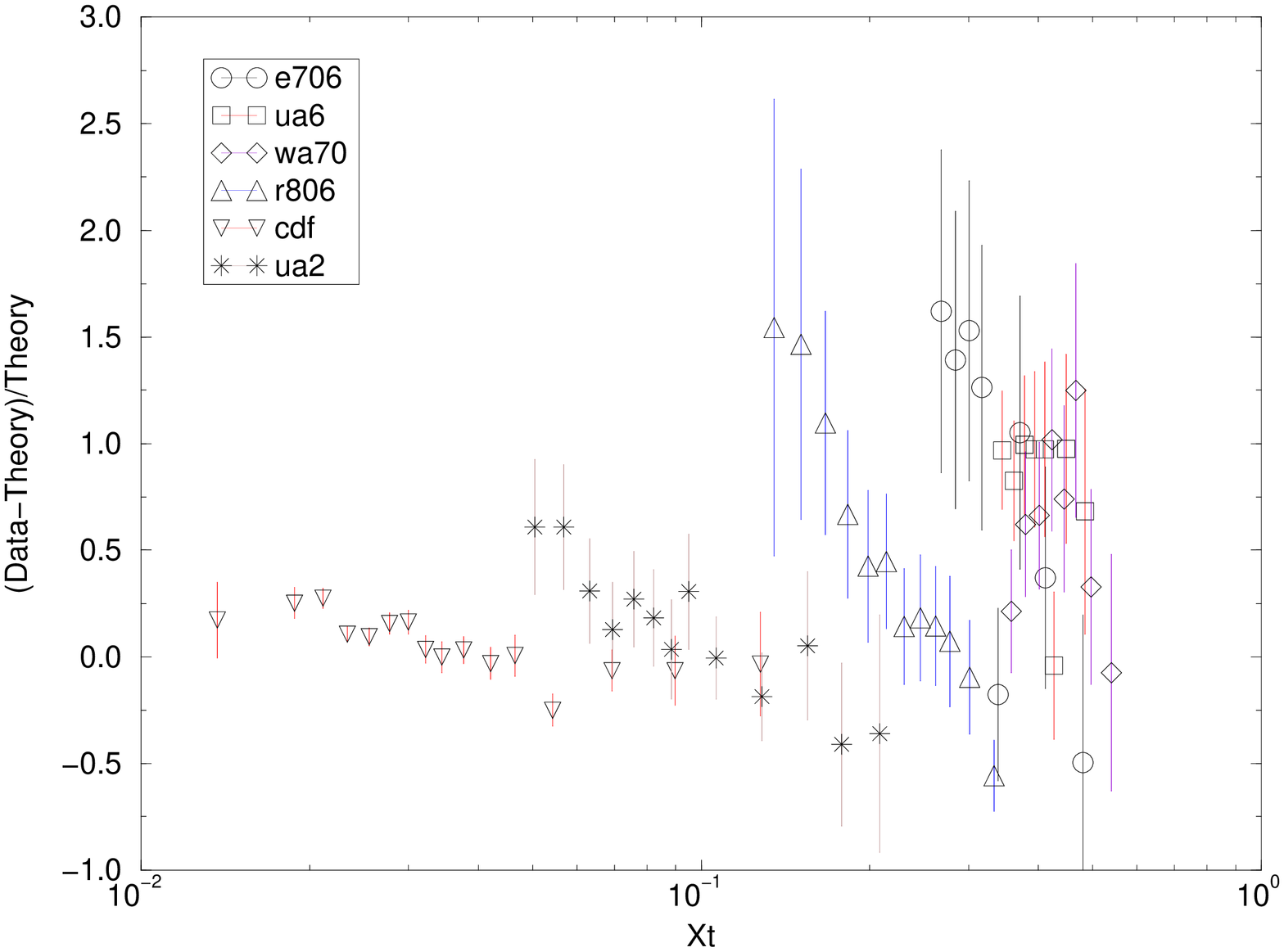}
\centering\includegraphics[angle=-90,width=.7\linewidth]{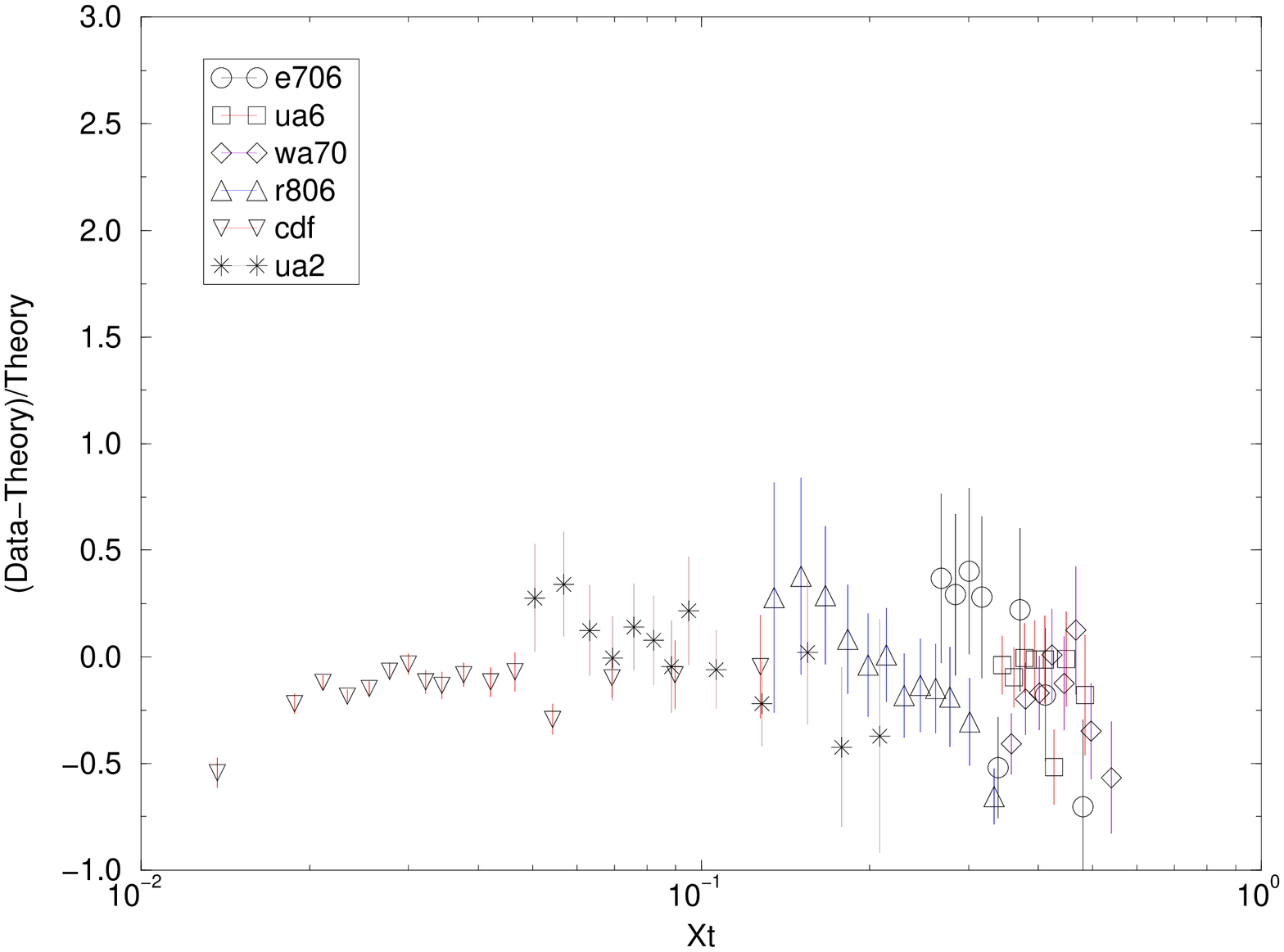}
\caption{Low $p_T$ direct photon spectra before (upper) and after (lower)
including the $k_T$ resummation effect.}
\label{direct}
\end{figure}

The $k_T$ resummation effect on the low $p_T$ spectra of the direct photon
production has been analyzed \cite{LL98}. The initial-state and final-state
radiations are constrained in the low $p_T$ region, where the $k_T$
resummation is necessary for improving the perturbation theory.
Figure~\ref{direct} shows the deviation, (Data -Theory)/Theory, of the
next-to-leading-order (NLO) pQCD predictions, obtained using the CTEQ4M PDFs
\cite{cteq4}, from the experimental data as
a function of $x_t=2p_T/\sqrt{s}$, $\sqrt{s}$ being the center-of-mass energy.
The deviation is huge as expected, especially at low $x_t$ of each
set of the data. After including the $k_T$ resummation
effect \cite{LL98}, it is clear that a significant improvement on the
agreement between the theoretical predictions and the data is achieved.
As to the intermediate- and high-$p_T$ regions of the direct photon production,
NLO pQCD works reasonably well in accommodating the data as shown in
\cite{Becher12}. The threshold resummation effect, which will be introduced
in the next subsection, is more relevant in these regions: it slightly improves
the consistency between predictions and the data \cite{Becher12}.

In the small $x$ region with $xp^+\sim k_T$, or $xp^+\sim 1/b$ in the
$b$ space, the two-scale case reduces to the single-scale one. In this region
contributions from gluonic partons dominate, so $\Phi$ represents the gluon
TMD below. The source of double logarithms, i.e., the integral containing
the anomalous dimension $\gamma_K$, is less important. Because only the soft
scale exists, one drops the hard kernel $G$, and keeps the soft kernel with an
ultraviolet cutoff. The right-hand side of Eq.~(\ref{meq}) becomes
\begin{eqnarray}
{\bar \Phi}(x,k_T)&=&
-ig^2N_c\int\frac{d^{4}l}{(2\pi)^4}\frac{{\hat n}\cdot v}{n\cdot l v\cdot l}
\left[\frac{\Theta(k_T^2-l_T^2)}{l^2}\Phi(x,k_T)\right.
\nonumber \\
& &\left.+2\pi i\delta(l^2)\phi(x,|{\bf k}_T+{\bf l}_T|)\right],
\label{kf1}
\end{eqnarray}
where the color factor $C_F$ has been replaced by $N_c$ for the gluon TMD.
The step function $\Theta$ introduces the ultraviolet cutoff on $l_T$ mentioned
above. To make variation in $x$ via variation in $p^+$, a fixed
parton momentum is assumed. Under this assumption, the momentum fraction $x$ is
proportional to $1/p^+$, and one has $p^+d\Phi/dp^+=-xd\Phi/dx\Phi$ \cite{L0}.
Performing the integrations over $l^+$ and $l^-$ in Eq.~(\ref{kf1}), the master
equation (\ref{meq}) reduces to the BFKL equation \cite{KMS},
\begin{eqnarray}
\frac{d\phi(x,k_T)}{d\ln(1/x)}=
{\bar \alpha}_s\int\frac{d^{2}l_T}{\pi l_T^2}
\left[\phi(x,|{\bf k}_T+{\bf l}_T|)
-\Theta(k_T^2-l_T^2)\phi(x,k_T)\right],
\label{bfkl}
\end{eqnarray}
with the coupling constant ${\bar \alpha}_s=N_c\alpha_s/\pi$.

A remarkable prediction of the above LO BFKL equation is that a
high-energy cross section increases with the center-of-mass energy,
\begin{eqnarray}
\sigma\approx\frac{1}{t}
\left(\frac{s}{t}\right)^{\omega_P-1},\label{llb}
\end{eqnarray}
with the momentum transfer squared $t$. It turns out that Eq.~(\ref{llb}),
with the Pomeron intercept $\omega_P-1=4{\bar \alpha}_s\ln 2$, violates the
Froissart (unitarity) bound $\sigma< {\rm const.}\times \ln^2 $ \cite{F61}.
The unsatisfactory prediction of the LO BFKL equation called for
the NLO corrections \cite{BFKLNLO}, which were, however, found to be dramatic
\cite{Thorne99}: the NLO effect is
nearly as large as the LO result for $x \sim 0.001$, and becomes dominant at
lower $x$. It even turns the derivative of the structure function $dF_L/d\ln Q^2$
negative below $x \sim 0.0001$.
That is, the perturbative solution is not at all
stable.

\section{Threshold Resummation and DGLAP Equation}

We then consider the soft approximation corresponding to
the $k_T$ ordering of real gluon emissions in a ladder diagram.
Assume that a parton without the transverse momentum,
carries $-{\bf l}_{1T}$ after emitting a gluon of longitudinal
momentum $l_1^+$ and transverse momentum ${\bf l}_{1T}$,
and then carries $-{\bf l}_{1T}-{\bf l}_{2T}$ after emitting a gluon of
longitudinal momentum $l_2^+$ and transverse momentum ${\bf l}_{2T}$.
In the kinematic configuration with $l_{2T}\gg l_{1T}$
and $l_2^+\sim l_1^+$, the final parton momentum can be approximated by
$-{\bf l}_{2T}-{\bf l}_{1T}\approx -{\bf l}_{2T}$, such that the loop
integral associated with the first gluon emission involves only $l_{1T}$,
and can be worked out straightforwardly, giving a logarithm.
The loop integral associated with the second gluon emission involves
only $l_{2T}$, and also gives a logarithm. Therefore, a ladder diagram
with $N$ rung gluons generates the logarithmic correction $(\alpha_s L)^N$
under the above $k_T$ ordering. In this case $\Phi$ is independent of $l_T$,
and we have the approximation for the real gluon emission in Eq.~(\ref{fsr})
\begin{equation}
\Phi(x+l^+/p^+,|{\bf k_T+\bf l_T}|)\approx
\Phi(x+l^+/p^+,k_T),
\label{nk}
\end{equation}
in which $x$ and $l^+/p^+$ are of the same order.
The dependence on $k_T$ can then be integrated out from both sides of
the master equation (\ref{meq}), and the TMD $\Phi$ reduces to the PDF $\phi$.
Similarly, the soft contribution $\bar\Phi_s$ in Eq.~(\ref{ssf}) reduces to
$\bar\phi_s$. The scale $k_T$ disappears, and the scale $(1-x)p^+$ is retained.

The Mellin transformation is employed to bring ${\bar\phi}_{s}$ from the
momentum fraction $x$ space to the moment $N$ space,
\begin{eqnarray}
{\bar\phi}_{s}(N)&=&\int_0^1 dxx^{N-1}{\bar\phi}_{s}(x),
\end{eqnarray}
under which the $l^+$ integration decouples.
In the large $x$ region $\phi$ involves two scales, the large
$xp^+\sim p^+$ from the hard kernel $G$ and the small
$(1-x)p^+\sim p^+/N$ from the soft kernel $K$.
To sum $\ln(1/N)$, we rewrite the derivative $p^+d\phi/dp^+$ as
\begin{equation}
p^+\frac{d\phi}{dp^+}=
\frac{p^+}{N}\frac{d\phi}{d (p^+/N)}.
\end{equation}
The solution of the master equation (\ref{meq}) then gives the threshold resummation,
\begin{eqnarray}
\phi(N)=\Delta_t(N)\phi_i
\label{pht}
\end{eqnarray}
with the exponential
\begin{eqnarray}
\Delta_t(N)=\exp\left[-2\int_{p^+/N}^{p^+}\frac{d p}{p}
\int_{p^+}^{p}\frac{d\mu}{\mu}
\gamma_{K}(\alpha_s(\mu))\right],
\label{fbt}
\end{eqnarray}
or its equivalent expression
\begin{eqnarray}
\Delta_t(N)=\exp\left[\int_{0}^{1}dz\frac{1-z^{N-1}}{1-z}
\int_{(1-z)^2}^{1}\frac{d\lambda}{\lambda}
\gamma_{K}(\alpha_s(\sqrt{\lambda}p^+))\right].
\end{eqnarray}
It has been investigated that Eq.~(\ref{fbt}) becomes reliable
as $N$ is about or greater than $O(10^2)$ at the Tevatron energy
\cite{Lai:1999ik}. Equation~(\ref{fbt}) is accurate up to next-to-leading
logarithms (NLL), so corrections to it appear at
next-to-next-to-leading logarithms (NNLL) and at powers of $1/N$.

An application of the threshold resummation is found in the analysis of the
top-quark pair production, which was performed at the NNLL
accuracy \cite{Beneke11}. It has been observed that
the threshold resummation effect enhances the NLO total cross section by
few percents, which make an impact on the determination of the top quark
mass.

In the intermediate $x$ region the two-scale case reduces to the
single-scale one because of $xp^+\sim (1-x)p^+$, and the source of double
logarithms is less important. Without the Mellin transformation,
the sum in Eq.~(\ref{ssf}), with the approximation
in Eq.~(\ref{nk}) being inserted, leads to the DGLAP equation \cite{L0},
\begin{eqnarray}
p^+\frac{d}{dp^+}\phi(x)
=\int_x^1 \frac{d\xi}{\xi}P(x/\xi)\phi(\xi)\;,
\label{con}
\end{eqnarray}
with the kernel
\begin{eqnarray}
P(z)=\frac{\alpha_s(p^+)}{\pi}C_F\frac{2}{(1-z)_+}\;,
\label{kgir}
\end{eqnarray}
where the variable change $\xi=x+l^+/p^+$ has been made. The argument
of $\alpha_s$, i.e, the factorization scale $\mu$,
has been set to the scale $xp^+\sim (1-x)p^+\sim O(p^+)$. Note that the
kernel $P$ differs from the splitting function $P_{qq}$
\begin{eqnarray}
P_{qq}^{(1)}(x)=C_F\left(\frac{1+x^2}{1-x}\right)_+,\label{splitq}
\end{eqnarray}
by the term $(z^2-1)/(1-z)_+$, which is finite in the $z\to 1$ limit.
The reason is that the real gluon emission was evaluated under the soft
approximation as deriving $P$, while it was calculated exactly as deriving
$P_{qq}$.

\begin{figure}
\centering\includegraphics[width=.7\linewidth]{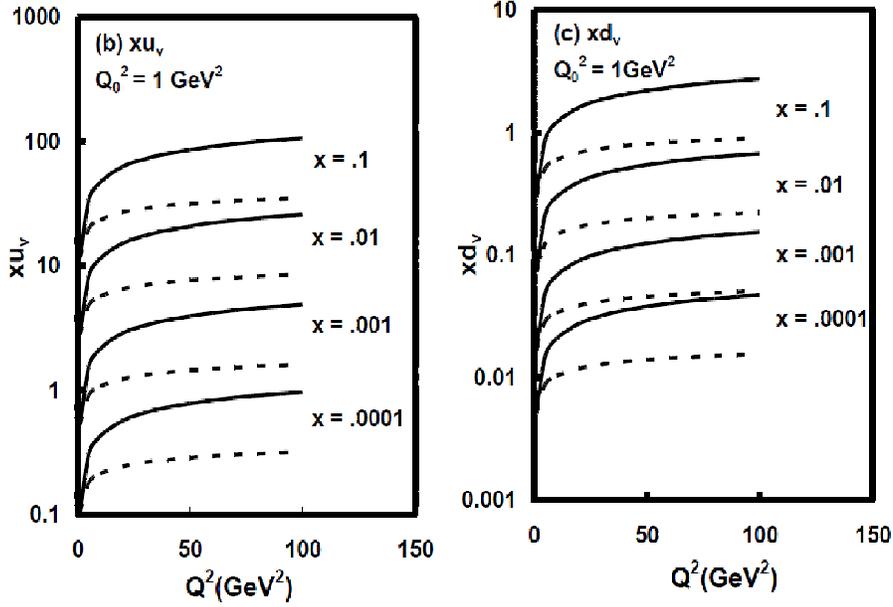}
\caption{$Q^2$ evolutions of the valence quark PDFs for some parameter values
in the DGLAP solutions (solid and dashed lines).
}
\label{dglap3}
\end{figure}

Gluon emissions cause the mixing between the quark and
gluon PDFs, giving the complete set of DGLAP equations with four splitting
functions
\begin{equation}\label{qs}
   \frac{\partial}{\partial \ln Q^2}\left( \begin{array}{c}
    \phi_q \\ \phi_g
   \end{array} \right)
   = \left( \begin{array}{cc}
    P_{qq} & P_{qg} \\ P_{gq} & P_{gg}
   \end{array} \right)\otimes
   \left( \begin{array}{c}
    \phi_q \\ \phi_g
   \end{array} \right).
\end{equation}
The evolution of the $u$-quark and $d$-quark PDFs in $Q^2$ predicted
by the LO DGLAP equation \cite{RS12} is shown in Fig.~\ref{dglap3},
where the inputs at the initial scale $Q_0=1$ GeV were taken from
MRST2001 \cite{MRST01}. It is observed that the valence quark PDFs increase
with $Q^2$ at small $x$, namely, they become broader with $Q^2$.

\section{Joint Resummation and CCFM Equation}

At last, a unified resummation formalism for large and intermediate
$x$ and a unified evolution equation for intermediate and small $x$ can be derived
by retaining the $l^+$ and $l_T$ dependencies of $\Phi$ in Eq.~(\ref{fsr}),
which corresponds to the so-called angular ordering.
In this case both the Fourier and Mellin transformations are applied
to Eq.~(\ref{fsr}), leading to
\begin{eqnarray}
{\bar\Phi}_s(N,b)=K(p^+/(N\mu),1/(b\mu),\alpha_s(\mu))\Phi(N,b)\;,
\end{eqnarray}
with the soft kernel \cite{Li99}
\begin{eqnarray}
K&=&-ig^2C_F\mu^\epsilon\int_0^1dz
\int\frac{d^{4-\epsilon}l}{(2\pi)^{4-\epsilon}}
\frac{{\hat n}\cdot v}{n\cdot l v\cdot l}
\left[\frac{\delta(1-z)}{l^2}\right.
\nonumber\\
& &\left.+2\pi i\delta(l^2)\delta\left(1-z-\frac{l^+}{p^+}\right)
z^{N-1}e^{i{\bf l}_T\cdot{\bf b}}\right]-\delta K,
\nonumber\\
&=&\frac{\alpha_s(\mu)}{\pi}C_F\left[\ln\frac{1}{b\mu}
-K_0\left(\frac{2\nu p^+b}{N}\right)\right],
\label{uk}
\end{eqnarray}
$K_0$ being the modified Bessel function. As $p^+b\gg N$, we have $K_0\to 0$,
and the soft scale inferred by the above expression approaches $1/b$ for the
$k_T$ resummation. As $N\gg p^+b$, we have $K_0\approx -\ln(\nu p^+b/N)$,
and the soft scale approaches $p^+/N$ for the threshold resummation.

\begin{figure}
\centering\includegraphics[width=.5\linewidth]{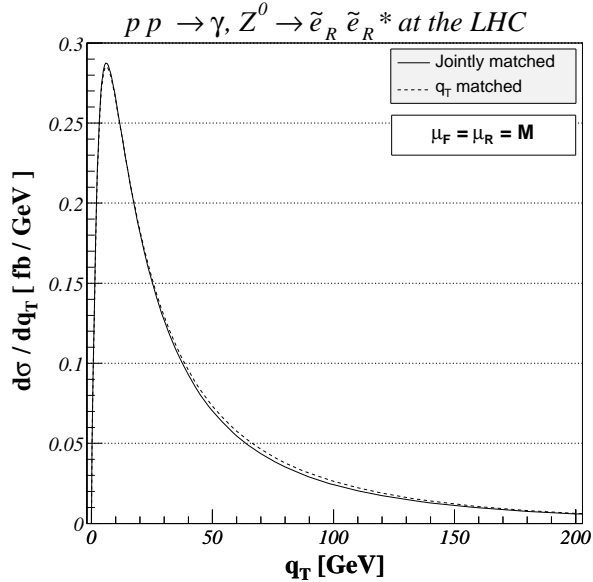}
\caption{Transverse-momentum distribution of selectron pairs at the LHC
in the framework of joint (full) and $k_T$ (dotted) resummations.
}
\label{joint1}
\end{figure}

Following the procedures similar to Eqs.~(\ref{dph})-(\ref{fb}),
we derive the joint resummation
\begin{equation}
\Phi(N,b)=\Delta_u(N,b)\Phi_i,
\end{equation}
with the exponential
\begin{equation}
\Delta_u(N,b)=
\exp\left[-2\int_{p^+\chi^{-1}(N,b)}^{p^+}\frac{d p}{p}
\int_{p^+\chi^{-1}(1,b)}^{p}\frac{d\mu}{\mu}
\gamma_{K}(\alpha_s(\mu))\right].
\label{fb3}
\end{equation}
The dimensionless function \cite{LSV00}
\begin{eqnarray}
\chi(N,b)=\left(N+\frac{p^+b}{2}\right)e^{\gamma_E},
\end{eqnarray}
is motivated by the limits discussed above. It is apparent that Eq.~(\ref{fb3})
reduces to Eq.~(\ref{fb}) and Eq.~(\ref{fbt}) in the $b\to\infty$ and
$N\to \infty$ limits, respectively. The effect from the joint resummation on
the $q_T$ spectra of selectron pairs produced at the LHC with $\sqrt{S}=14$ TeV
has been investigated in \cite{Bozzi07}.
It is seen in Fig.~\ref{joint1} that the joint and $k_T$ resumations exhibit a
similar behavior in the small-$q_T$ region as expected, but the jointly-resummed
cross section is about 5\%-10\% lower than the $k_T$-resummed cross section in
the range 50 GeV $< q_T <$ 100 GeV.

In the intermediate and small $x$ regions, it is not necessary to resum the
double logarithms $\ln^2(1/N)$. After extracting the $k_T$ resummation, the
remaining single-logarithmic summation corresponds to a unification of the
DGLAP and BFKL equations, since both the $l^+$ and $l_T$ dependencies have
been retained. The function $\Phi(x+l^+/p^+,b)$ in Eq.~(\ref{fsr}) is reexpressed,
after the Fourier transformation, as
\begin{eqnarray}
& &\Phi(x+l^+/p^+,b)=\Theta((1-x)p^+-l^+)\Phi(x,b)
\nonumber\\
& &\hspace{1.5cm} +[\Phi(x+l^+/p^+,b)-\Theta((1-x)p^+-l^+)
\Phi(x,b)].
\label{fre}
\end{eqnarray}
The contribution from the first term is combined with the first term in
Eq.~(\ref{fsr}), giving the soft kernel $K$ for the $k_T$ resummation.
The second term in Eq.~(\ref{fre}) contributes
\begin{eqnarray}
-iN_cg^2\int\frac{d^4l}{(2\pi)^4}
\frac{{\hat n}\cdot v}{n\cdot l v\cdot l}
2\pi i\delta(l^2)e^{i{\bf l}_T\cdot {\bf b}}
[\Phi(x+l^+/p^+,b)-\Theta((1-x)p^+-l^+)\Phi(x,b)],
\label{fs2}
\end{eqnarray}
which will generate the splitting function below. The color factor has
been replaced by $N_c$, since the gluon TMD is considered here.

The master equation (\ref{meq}) then becomes
\begin{eqnarray}
p^+\frac{d}{dp^+}\Phi(x,b)&=&-2\left[\int_{1/b}^{xp^+}
\frac{d\mu}{\mu}\gamma_K(\alpha_s(\mu))
-{\bar \alpha}_s(xp^+)\ln(p^+b)\right]\Phi(x,b)
\nonumber\\
& &+2{\bar\alpha}_s(xp^+)\int_x^1 dz P_{gg}(z)\Phi(x/z,b),
\label{ue2}
\end{eqnarray}
with the splitting function
\begin{equation}
P_{gg}=\left[\frac{1}{(1-z)_+}+\frac{1}{z}-2+z(1-z)\right],
\label{pgg}
\end{equation}
obtained from Eq.~(\ref{fs2}). The term $-2+z(1-z)$ finite as
$z\to 0$ and $z\to 1$ has been added. The exponential $\Delta$ is extracted
from the $k_T$ resummation,
\begin{eqnarray}
\Delta(x,b,Q_0)
=\exp\left(-2\int_{xQ_0}^{xp^+}\frac{dp}{p}
\left[\int_{1/b}^{p}
\frac{d\mu}{\mu}\gamma_K(\alpha_s(\mu))
-{\bar \alpha}_s(p)\ln\frac{p b}{x}\right]\right),
\label{del}
\end{eqnarray}
$Q_0$ being an arbitrary low energy scale. It is trivial to justify by
substitution that the solution is given by
\begin{eqnarray}
\Phi(x,b)&=&\Delta(x,b,Q_0)\Phi_i
\nonumber\\
& &+2\int_x^1 dz\int_{Q_0}^{p^+}\frac{d\mu}{\mu}
{\bar\alpha}_s(x\mu)\Delta_k(x,b)P_{gg}(z)\Phi(x/z,b),
\label{nunif}
\end{eqnarray}
which can be regarded as a modified version of the CCFM equation \cite{CCFM}.

\section{JET MASS DISTRIBUTION}

Jets, abundantly produced at colliders \cite{SW77}, carry information of
hard scattering and parent particles, which is crucial for particle
identification and new physics search. Study of jet physics usually relies
on event generators, which, however, suffer ambiguity from parameter
tuning. Hence, we are motivated to establish an alternative approach free of
the ambiguity. I will demonstrate that jet dynamics can be explored and jet
properties can be predicted in the resummation formalism with the Wilson
lines off the light cone.

We start from the dijet production in the $e^-e^+$ annihilation, which is part of
its total cross section. The physical dijet final state contains two jet cones of
half angle $\delta$ and isotropic soft gluons within the energy resolution
$\epsilon Q$, $Q$ being the $e^-e^+$ invariant mass. With the
constrained phase space for real gluons, the infrared cancellation is not complete,
and logarithmic enhancement appears. The explicit NLO calculations imply that the
isotropic soft gluons give a contribution proportional to $2\ln^2(2\epsilon Q/\mu)-\pi^2/6$,
the collinear gluons in the cones with energy higher than the resolution give
$-3\ln(Q\delta/\mu)-2\ln^2(2\epsilon)-4\ln(Q\delta/\mu)\ln(2\epsilon)+17/4-\pi^2/3$,
and the virtual corrections contribute $-2\ln^2(Q/\mu)+3\ln(Q/\mu)-7/4+\pi^2/6$. The total
NLO corrections indicate that the dijet cross section is infrared finite, but
logarithmically enhanced \cite{S95}:
\begin{eqnarray}
3\ln\delta+4\ln\delta\ln(2\epsilon)+\frac{\pi^2}{3}-\frac{5}{2},\label{dijet}
\end{eqnarray}
where the double logarithm $\ln\delta\ln(2\epsilon)$ is attributed to the
overlap of the collinear and soft logarithms.

We then explain the factorization of a jet from DIS, whose production is
expected to be enhanced by collinear dynamics as indicated by
Eq.~(\ref{dijet}). A jet is formed, as the gluon emitted by the
initial-state or final-state quark is collimated to the final-state quark.
The restricted phase space of the final-state quark and the gluon in a
small angular separation renders an incomplete cancellation between the
virtual and real corrections. In this kinematic configuration the
initial-state quark propagator can be eikonalized, such that collinear gluons
are detached from the initial-state quark and absorbed into a jet function.
To all orders, the collinear gluons are collected by the Wilson link
with an arbitrary vector $n$. The collinear gluon emitted by the final-state
quark can be factorized into the jet function straightforwardly by
applying the Fierz transformation. A more sophisticated factorization
formula for the jet production in the DIS is then written as a convolution
of a hard kernel $H$ with a PDF and a jet function $J$.
$H$ denotes the contribution with the collinear pieces for the
initial and final states being subtracted.

The light-quark and gluon jet functions are defined by \cite{Almeida:2008tp}
\begin{eqnarray}
J_q(M_J^2,P_T,\nu^2,R,\mu^2)&=&\frac{(2\pi)^3}
{2\sqrt{2}(P_J^0)^2N_c}\sum_{N_J}Tr\left\{\not\xi\langle
0|q(0)W^{(\bar q)\dagger}|N_J\rangle\langle N_J|W^{(\bar q)}
\bar q(0)|0\rangle\right\}\nonumber\\
& &\times\delta(M_J^2-\hat M_J^2(N_J,R))\delta^{(2)}(\hat e-\hat
e(N_J))\delta(P_J^0-\omega(N_J)),
\nonumber \\
J_g(M_J^2,P_T,\nu^2,R,\mu^2)&=&\frac{(2\pi)^3}
{2(P_J^0)^3N_c}\sum_{N_J}\langle
0|\xi_\sigma F^{\sigma\nu}(0)W^{(g)\dagger}
|N_J\rangle\langle N_J|W^{(g)}
F_\nu^\rho(0)\xi_\rho|0\rangle\nonumber\\
& &\times\delta(M_J^2-\hat M_J^2(N_J,R))\delta^{(2)}(\hat e-\hat
e(N_J))\delta(P_J^0-\omega(N_J)),\label{jet1}
\end{eqnarray}
where $|N_J\rangle$ denotes the final state with $N_J$ particles
within the cone of size $R$ centered in the direction of the unit
vector $\hat e$, $\hat M_J(N_J,R)$ ($\omega(N_J)$) is the invariant mass
(total energy) of all $N_J$ particles, and $\mu$ is the factorization scale.
The above jet functions absorb the collinear divergences from all-order
radiations associated with the energetic light jet of momentum
$P_J^\mu=P_J^0 v^\mu$, in which $P_J^0$ is the jet energy, and
the vector $v$ is given by $v^\mu=(1,\beta,0,0)$ with
$\beta=\sqrt{1-(M_J/P_J^0)^2}$. $\xi^\mu=(1,-1,0,0)$ is a vector on the
light cone. The coefficients in Eq.~(\ref{jet1}) have been chosen, such
that the LO jet functions are equal to $\delta(M_J^2)$ in a perturbative
expansion.

Underlying events include everything but hard scattering, such as
initial-state radiation, final-state radiation, and multiple parton
interaction (MPI). The Wilson lines in Eq.~(\ref{jet1}) have collected
gluons radiated from both initial states and other final states in a
scattering process, and collimated to the light-particle jets. Gluon
exchanges between the quark fields $q$ (or the gluon fields $F^{\sigma\nu}$
and $F_\nu^\rho$) correspond to the final-state radiations. Both the
initial-state and final-state radiations are leading-power effects
in the factorization theorem, and have been included in the jet
function definition. A chance of involving more partons in hard scattering
is low, so the contribution from MPI is regarded as being subleading-power.
This contribution should be excluded from data, but it is certainly
difficult to achieve in experiments. Nevertheless, it still makes sense
to compare predictions for jet observables based on Eq.~(\ref{jet1})
at the current leading-power accuracy with experimental data. At last,
pile-up events must be removed in experiments \cite{SS13}, since they
cannot be handled theoretically so far.

\begin{figure}
\centering\includegraphics[width=.6\linewidth]{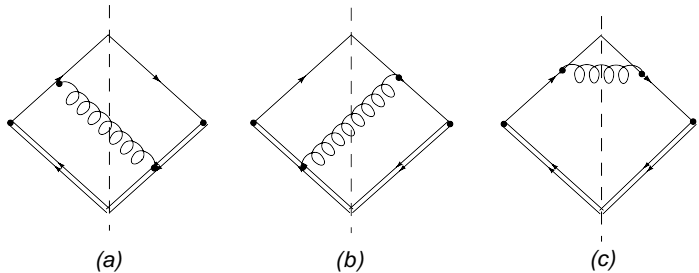}
\caption{Some NLO real corrections to the quark jet function.}
\label{quark}
\end{figure}

\begin{figure}
\centering\includegraphics[width=.5\linewidth]{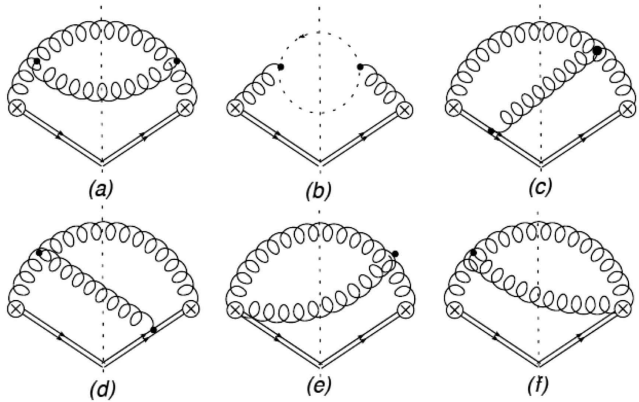}
\caption{Some NLO real corrections to the gluon jet function, where the
dashed line represents a ghost field.}
\label{gluon}
\end{figure}

\begin{figure}
\centering\includegraphics[width=.8\linewidth]{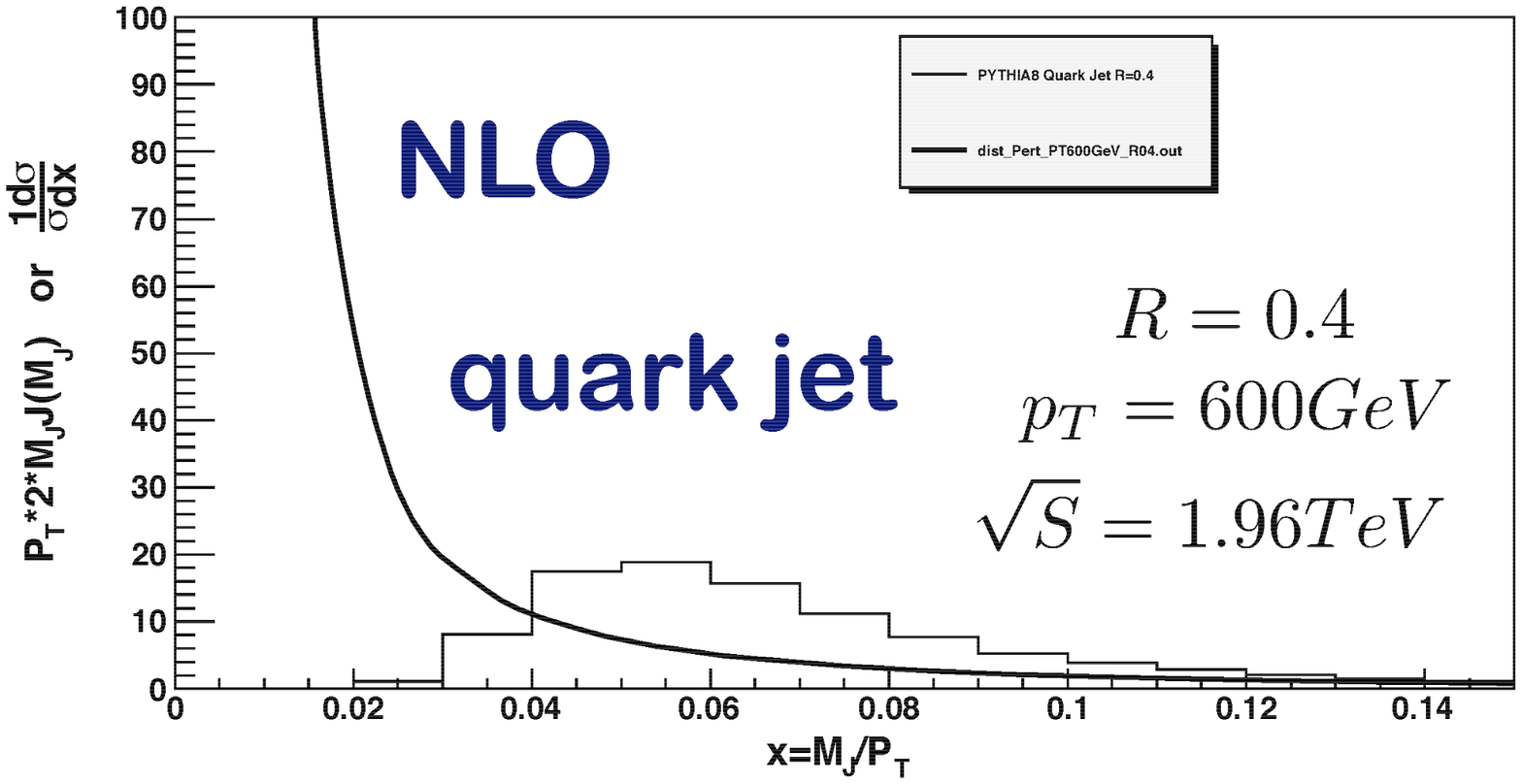}
\caption{Jet mass distribution at NLO.}
\label{jet4}
\end{figure}

The NLO diagrams for the light-quark and gluon jet functions are displayed in
Figs.~\ref{quark} and \ref{gluon}, respectively. Evaluating the jet functions
up to NLO, a divergence, compared to PYTHIA predictions, is observed at small
jet invariant mass $M_J$ as
shown in Fig.~\ref{jet4}, that implies the nonperturbtive nature of the
jet functions. The total NLO corrections in Mellin space indicate the
existence of double logarithms, which demand the implementation of the
resummation technique. Both the angular and energy resolutions are related to
the jet mass: when $M_J$ is not zero, particles in a jet cannot be completely
collimated, and the jet must have finite minimal energy. This accounts for
the source of the double logarithms. Recall that low $p_T$ spectra of
direct photons, dominated by soft and collinear radiations, are treated
by the $k_T$ resummation. The jet invariant mass is attributed to soft and
collinear radiations, so the mass distribution can also be derived in the
resummation formalism.

\begin{figure}[!htb]
\centering\includegraphics[width=0.7\textwidth]{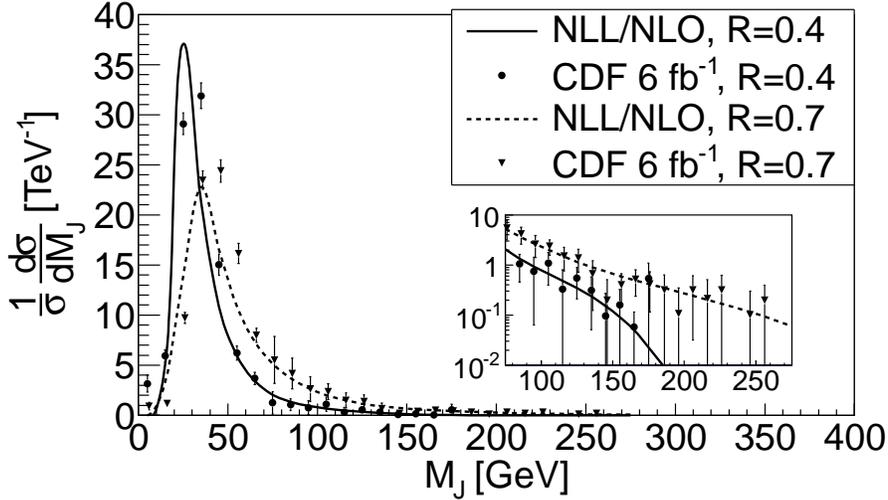}
\caption{Comparison of resummation predictions for the jet mass
distributions to Tevatron CDF data with the kinematic cuts $P_T>400$
GeV and $0.1<|Y|<0.7$ at $R=0.4$ and $R=0.7$. The inset shows the
detailed comparison in large jet mass region.} \label{CONVP1}
\end{figure}

Varying the Wilson line direction $n$, we derive the differential equation for the
light-quark jet function \cite{Li:2011hy}
\begin{eqnarray}
-\frac{n^2}{v\cdot n}v_{\alpha}\frac{d}{dn_\alpha}{
J}_q(M_J^2,P_T,\nu^2,R,\mu^2)=2(K+G)\otimes {J}_q(M_J^2,P_T,\nu^2,R,\mu^2). \label{cr2}
\end{eqnarray}
The above equation implies that the soft gluons in $K$ are associated
with the jet function $J$, a prescription consistent with the anti-$k_T$
algorithm. The strategy is to evolve $n$, i.e.,  $\nu^2$ from a low
value $\nu_{\rm in}^2\sim O(1/N)$ to a large value
$\nu_{\rm fi}^2\sim O(1)$. The former defines the initial condition of
the jet function, which can be evaluated up to a fixed order, because
of the vanishing of the logarithm $\ln(\nu^2 N)$. The latter reproduces
all important logarithms in the jet function, such that the solution of
Eq.~(\ref{cr2}) collects their resummation. One then convolutes the
light-quark and gluon jet functions with the constituent cross sections of
LO partonic dijet processes at the Tevatron and the PDF CTEQ6L \cite{Pumplin:2002vw}.
The resummation predictions for the jet mass distributions at $R=0.4$ and
$R=0.7$ are compared to the Tevatron CDF data \cite{Aaltonen:2011pg} in
Fig.~\ref{CONVP1} \cite{Li:2011hy} with the kinematic cuts $P_T>400$ GeV and the
rapidity interval $0.1<|Y|<0.7$. The abbreviation NLL refers to the accuracy of
the resummation at next-to-leading logarithm, and NLO to the accuracy of the
initial condition of the jet function solved from Eq.~(\ref{cr2}). The consistency
of the resummation results with the CDF data is satisfactory.

\section{JET ENERGY PROFILE}

It is known that a top quark produced almost at rest at the Tevatron
can be identified by measuring isolated jets from its decay.
However, this strategy does not work for identifying a
highly-boosted top quark produced at the LHC. It has been observed
that an ordinary high-energy QCD jet
\cite{Skiba:2007fw,Holdom:2007ap} can have an invariant mass close
to the top quark mass. A highly-boosted top quark, producing only
a single jet, is then difficult to be distinguished from a QCD jet.
This difficulty also appears in the identification of a
highly-boosted new-physics resonance decaying into standard-model
particles, or Higgs boson decaying into a bottom-quark pair. Hence,
additional information needs to be extracted from jet internal
structures in order to improve the jet identification at the LHC.
The quantity, called planar flow \cite{Almeida:2008yp}, has been
proposed for this purpose, which utilizes the geometrical shape of a
jet: a QCD jet with large invariant mass mainly involves one-to-two
splitting, so it leaves a linear energy deposition in a detector. A
top-quark jet, proceeding with a weak decay, mainly involves
one-to-three splitting, so it leaves a planar energy deposition.
Measuring this additional information, it has been shown with event
generators that the top-quark identification can be improved to some
extent. Investigations on various observables associated with jet
substructures are usually done using event generators. For a review on recent
theoretical progress and the latest experimental results in jet
substructures, see \cite{Altheimer:2012mn}.

Here I focus on a jet substructure, called the energy profile,
and explain how to calculate it in the resummation formalism \cite{Li:2011hy}.
This quantity describes the energy fraction accumulated
in the cone of size $r$ within a jet cone $R$, i.e., $r<R$. Its
explicit definition is given by \cite{Acosta:2005ix}
\begin{equation}
\Psi(r)=\frac{1}{N_{J}}\sum_{J}
\frac{\sum_{r_i<r,i\in {J}}P_{Ti}}{\sum_{r_i<R, i\in {J}}P_{Ti}},\label{pro}
\end{equation}
with the normalization $\Psi(R)=1$, where $P_{Ti}$ is the transverse
momentum carried by particle $i$ in the jet $J$, and $r_i<r$
$(r_i<R)$ means the flow of particle $i$ into the jet cone $r$
$(R)$. Different types of jets are expected to exhibit different
energy profiles. For example, a light-quark jet is narrower than a
gluon jet; that is, energy is accumulated faster with $r$ in a
light-quark jet than in a gluon jet. A heavy-particle jet certainly
has a distinct energy profile, which can be used for its identification.
The importance of higher-order corrections and
their resummation for studying a jet energy profile have been first
emphasized in \cite{Seymour:1997kj}. Another approach based on the
soft-collinear effective theory and its application to jet
production at an electron-positron collider can be found in Refs.
\cite{Ellis:2010rwa,Kelley:2011tj,Kelley:2011aa}.

We first define the jet energy
functions $J^E_f(M_J^2,P_T,\nu^2,R,r)$ with $f=q(g)$ denoting the
light-quark (gluon), which describe the energy accumulation
within the cone of size $r<R$. The definition is chosen, such that
$J^{E(0)}_f=P_T\delta(M_J^2)$ at LO. The Feynman rules for $J^E_f$
are similar to those for the jet functions $J_f$ at each order of
$\alpha_s$, except that a sum of the step functions
$\sum_ik_i^0\Theta(r-\theta_i)$  is inserted, where $k_i^0$
($\theta_i$) is the energy (the angle with respect to the jet axis)
of particle $i$. For example, the jet energy
functions $J^E_f$ are expressed, at NLO, as
\begin{eqnarray}
J_q^{E(1)}(M_J^2,P_T,\nu^2,R,r,\mu^2)&=&\frac{(2\pi)^3}
{2\sqrt{2}(P_J^0)^2N_c}\sum_{\sigma,\lambda}
\int\frac{d^3p}{(2\pi)^3 2p^0}\frac{d^3k}{(2\pi)^3 2k^0}\nonumber\\
& &\times [p^0\Theta(r-\theta_p)+k^0\Theta(r-\theta_k)]
\nonumber \\
&& \times{\rm Tr}\left\{\not\xi\langle 0|q(0)W^{(\bar
q)\dagger}|p,\sigma;k,\lambda\rangle \langle
k,\lambda;p,\sigma|W^{(\bar q)} \bar
q(0)|0\rangle\right\}\nonumber\\
& &\times\delta(M_J^2-(p+k)^2)\delta^{(2)}(\hat e-\hat
e_{\bf{p}+\bf{k}})\delta(P_J^0-p^0-k^0), \nonumber \\
J_g^{E(1)}(M_J^2,P_T,\nu^2,R,r,\mu^2)&=&\frac{(2\pi)^3}
{2(P_J^0)^3N_c}\sum_{\sigma,\lambda} \int\frac{d^3p}{(2\pi)^3
2p^0}\frac{d^3k}{(2\pi)^3 2k^0}\nonumber\\
& &\times [p^0\Theta(r-\theta_p)+k^0\Theta(r-\theta_k)]
\nonumber \\
&& \times \langle 0|\xi_\sigma
F^{\sigma\nu}(0)W^{(g)\dagger}
|p,\sigma;k,\lambda\rangle\langle
k,\lambda;p,\sigma|W^{(g)}
F_\nu^\rho(0)\xi_\rho|0\rangle\nonumber\\
& &\times\delta(M_J^2-(p+k)^2)\delta^{(2)}(\hat e-\hat
e_{\bf{p}+\bf{k}})\delta(P_J^0-p^0-k^0),
\label{jetENLO1}
\end{eqnarray}
where the expansion of the Wilson links in $\alpha_s$ is understood.
The factorization scale is set to $\mu=P_T$ to remove the associated
logarithms, so its dependence will be suppressed below.

\begin{figure}[!htb]
\centering\includegraphics[width=0.5\textwidth]{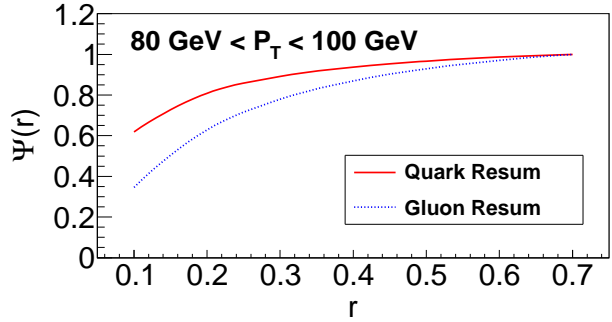}
\caption{Resummation predictions for the energy profiles of the
light-quark (solid curve) and gluon (dotted curve) jets with
$\sqrt{s}=7$ TeV and 80 GeV $< P_T<$ 100 GeV.} \label{COMP}
\end{figure}

The Mellin-transformed jet energy function ${\bar J}_q^E$ obeys a
similar differential equation \cite{Li:2011hy}
\begin{eqnarray}
-\frac{n^2}{v\cdot n}v_{\alpha}\frac{d}{dn_\alpha}{\bar
J}_q^E(N=1,P_T,\nu^2,R,r)=2({\bar K}+ G){\bar
J}_q^E(N=1,P_T,\nu^2,R,r), \label{er2}
\end{eqnarray}
which can be solved simply. Inserting the solutions to Eq.~(\ref{er2})
into Eq.~(\ref{pro}), the jet energy profile is derived. Note that a
jet energy profile with $N=1$ is not sensitive to
the nonperturbative contribution, so the predictions are free of the
nonperturbative parameter dependence, in contrast to the case of
the jet invariant mass distribution. It has been found
that the light-quark jet has a narrower energy profile than the
gluon jet, as exhibited in Fig.~\ref{COMP} for $\sqrt{s}=7$ TeV and
the interval $80$ GeV $< P_T<$ $100$ GeV of the jet transverse
momentum. The broader distribution of the gluon jet results from
stronger radiations caused by the larger color factor $C_A=3$,
compared to $C_F=4/3$ for a light-quark jet.

\begin{figure}[!htb]
\includegraphics[width=0.32\textwidth]{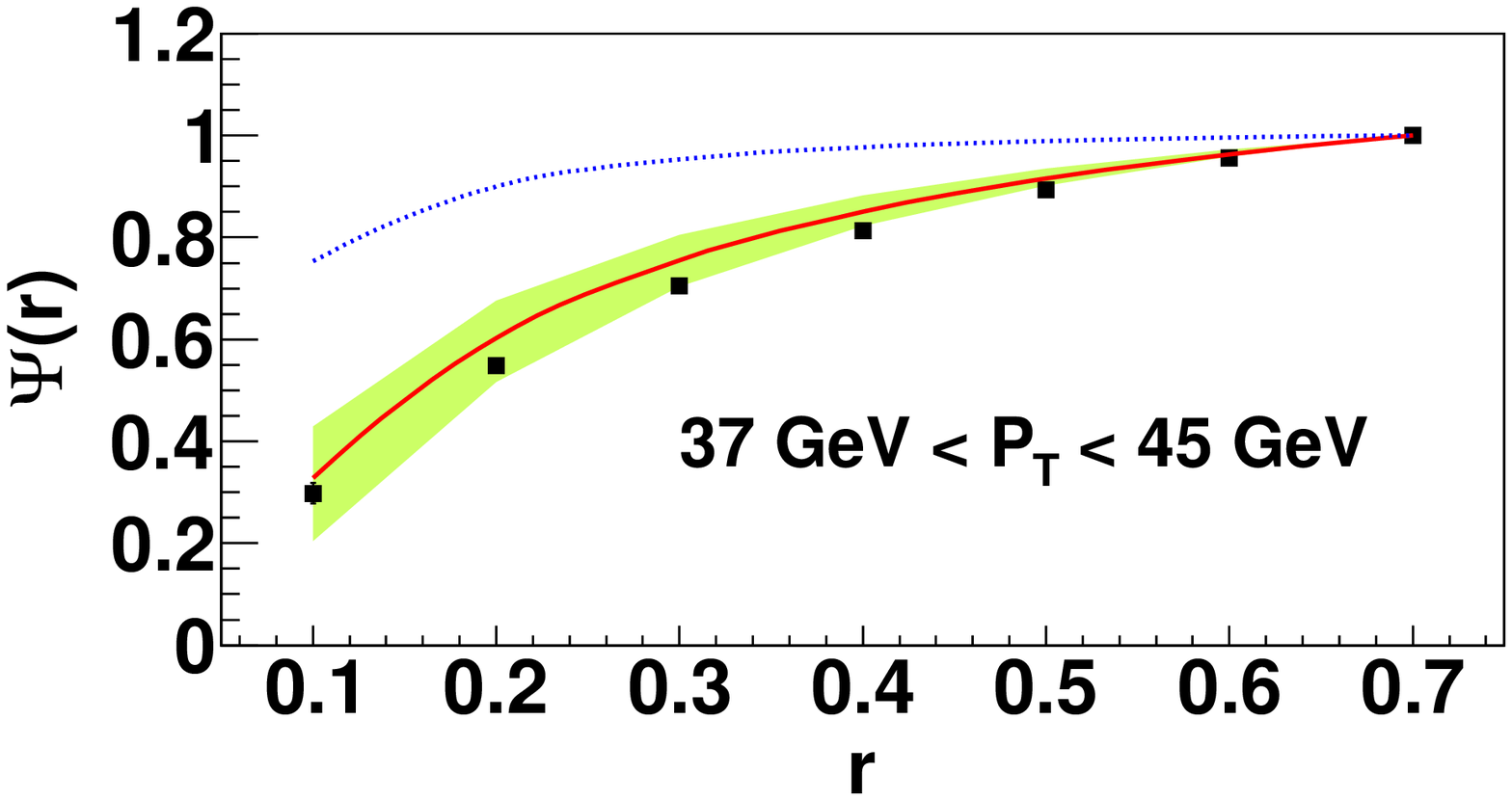}
\includegraphics[width=0.32\textwidth]{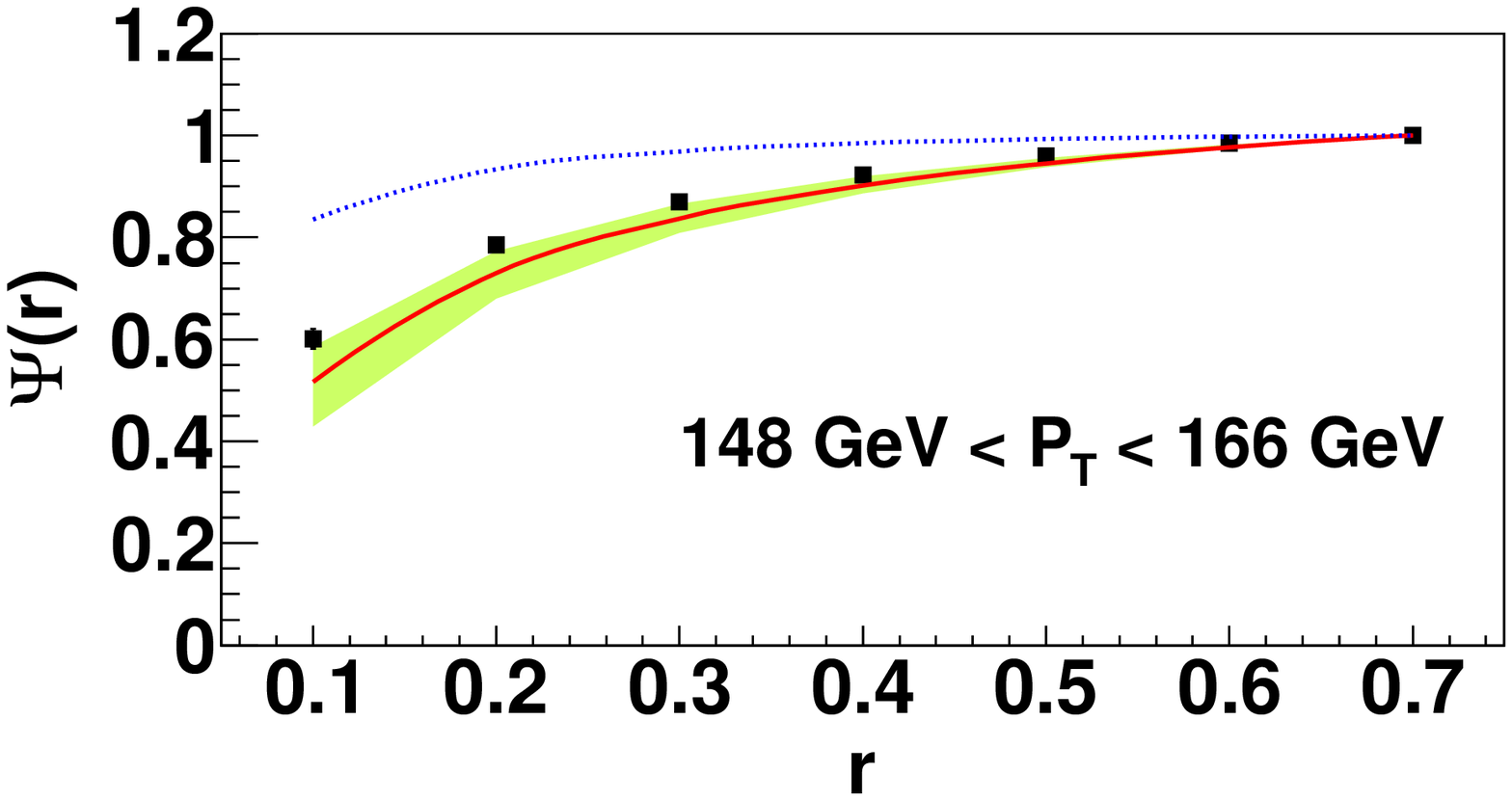}
\includegraphics[width=0.32\textwidth]{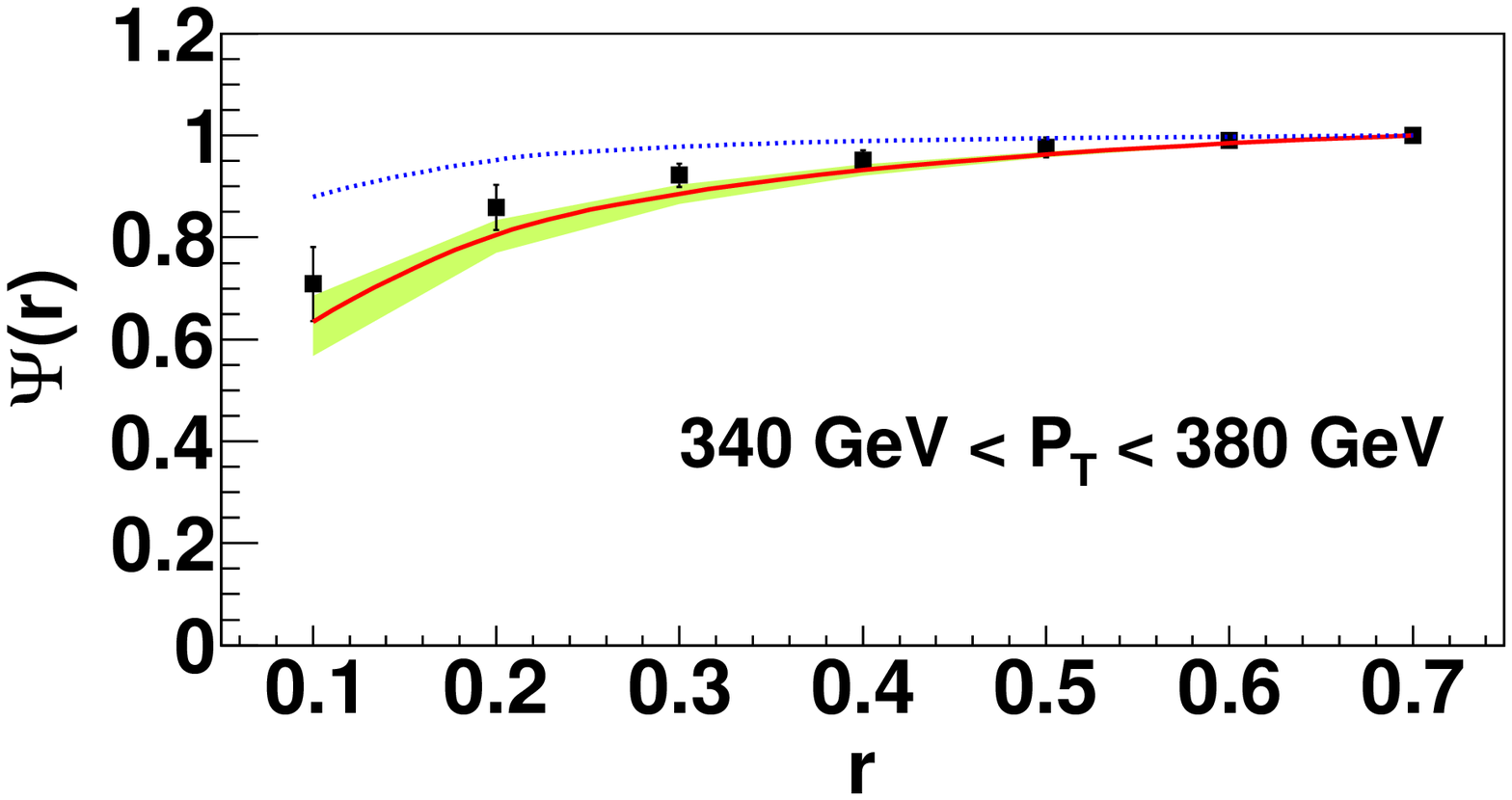}
\caption{Comparison of resummation predictions for the jet energy
profiles with $R=0.7$ to Tevatron CDF data in various $P_T$ intervals. The NLO
predictions denoted by the dotted curves are also displayed.}
\label{CDFJE}
\end{figure}

\begin{figure}[!htb]
\includegraphics[width=0.32\textwidth]{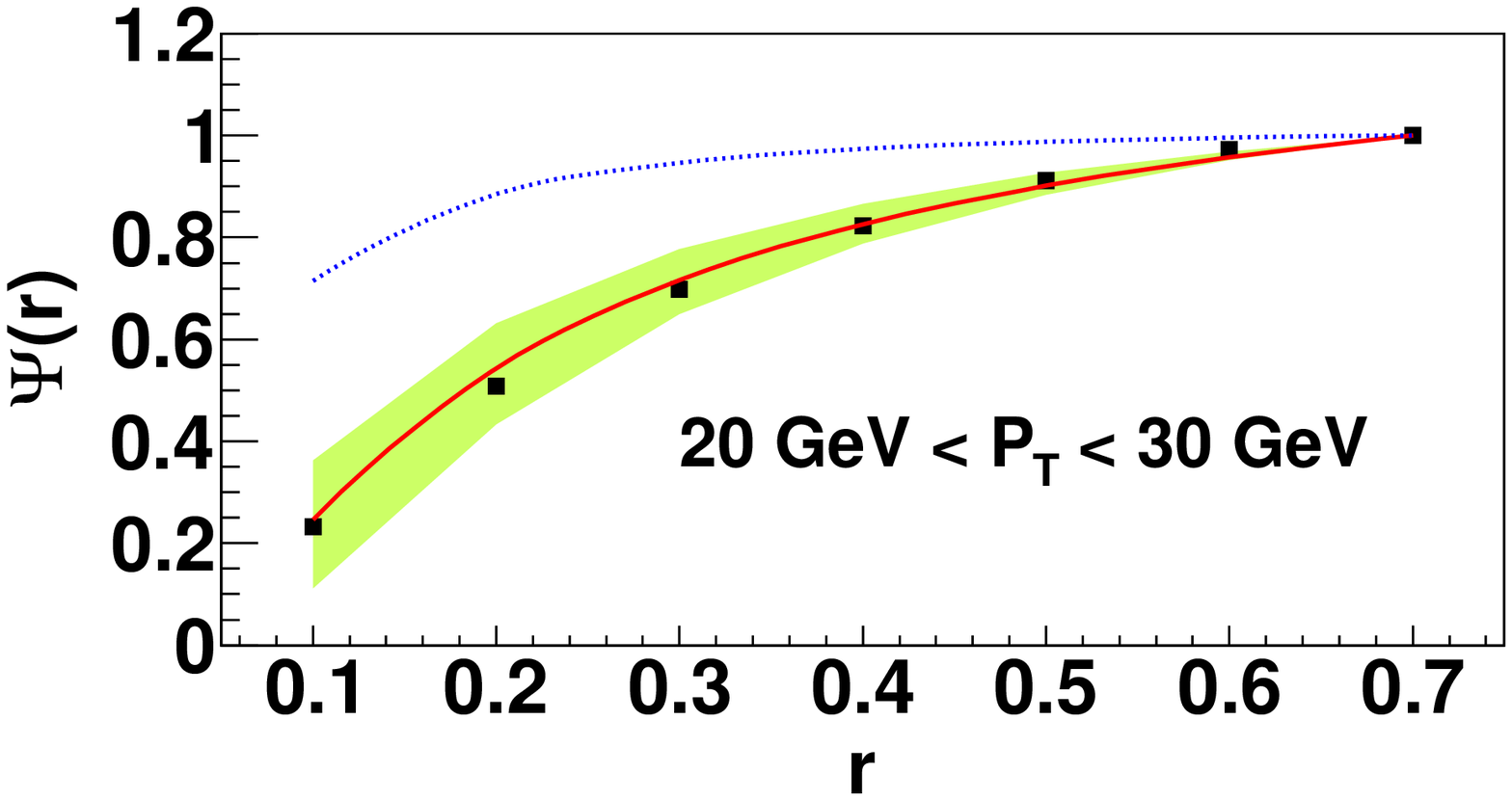}
\includegraphics[width=0.32\textwidth]{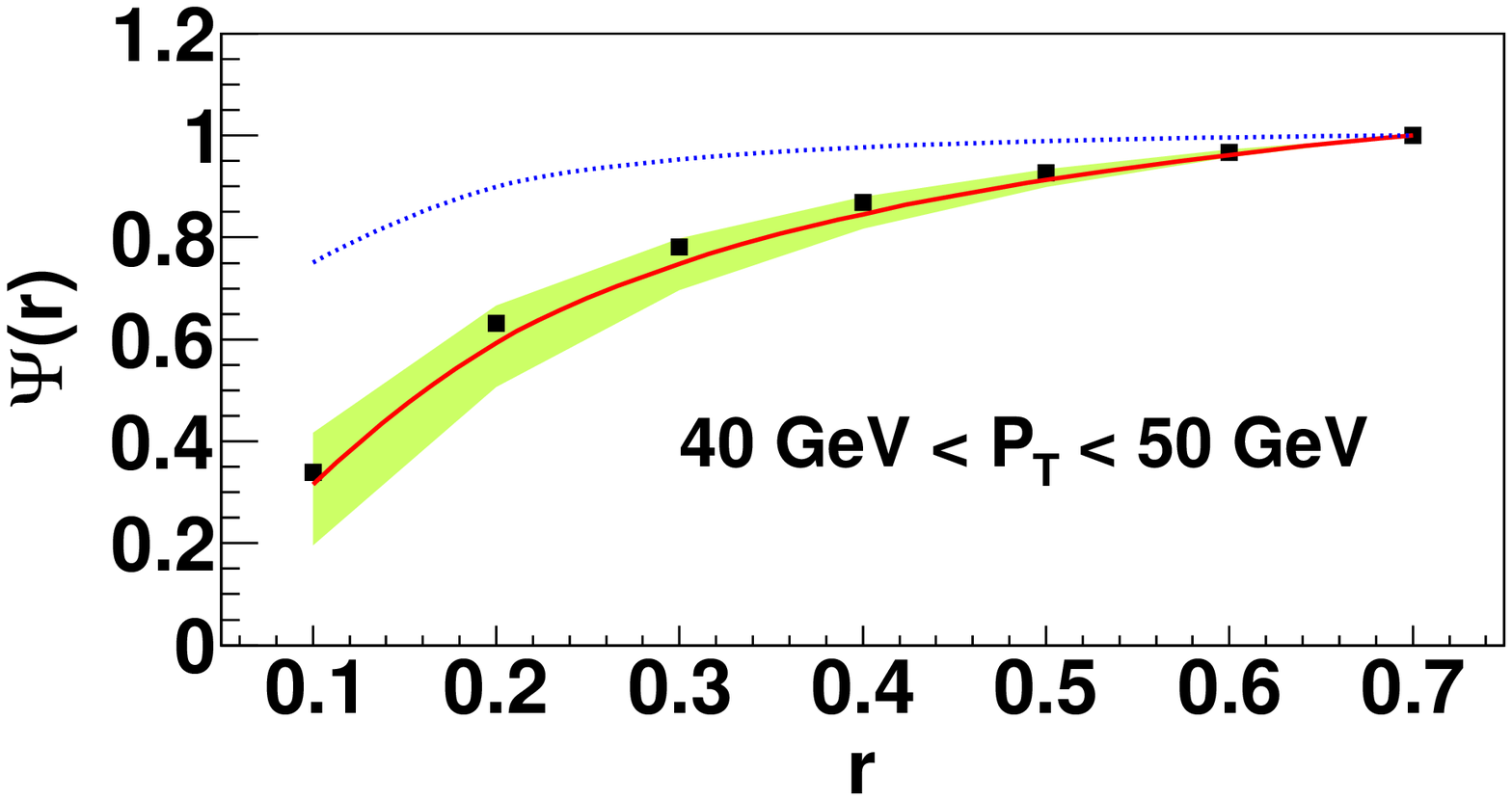}
\includegraphics[width=0.32\textwidth]{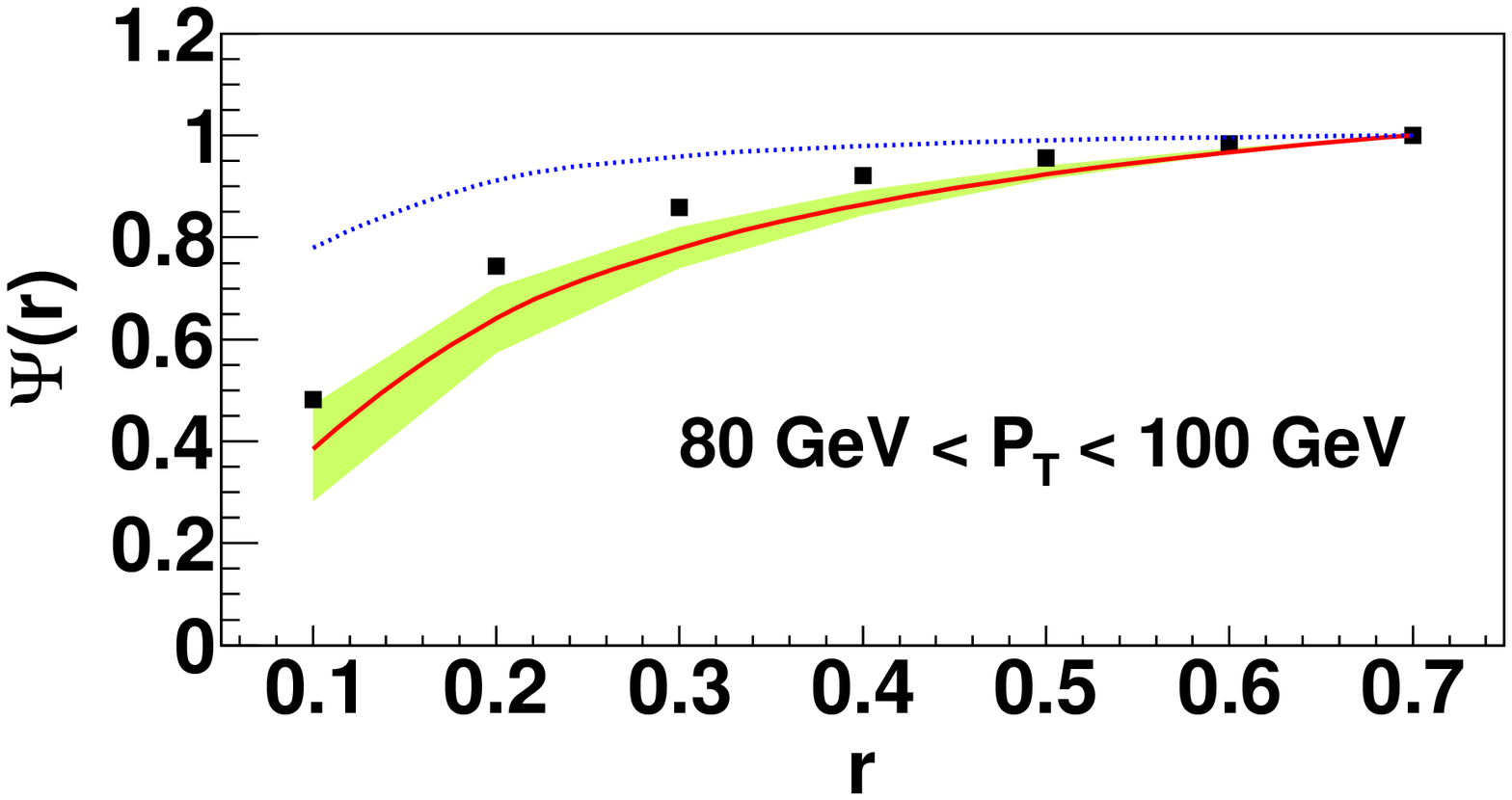}
\caption{Resummation predictions for the
jet energy profiles with $R=0.7$ compared to LHC CMS data
in various $P_T$ intervals. The NLO
predictions denoted by the dotted curves are also displayed.}
\label{CMSJE}
\end{figure}

One then convolutes the light-quark and gluon jet energy functions
with the constituent cross sections of the LO partonic subprocess
and CTEQ6L PDFs \cite{Pumplin:2002vw} at certain collider energy.
The predictions are directly compared with
the Tevatron CDF data \cite{Acosta:2005ix} as shown in Fig.~\ref{CDFJE}.
It is evident that the resummation predictions agree well
with the data in all $P_T$ intervals. The NLO predictions derived from
$\bar J_f^{E(1)}(1,P_T,\nu_{\rm fi}^2, R, r)$ are also displayed for
comparison, which obviously overshoot the data. The resummation
predictions for the jet energy profiles are compared with the LHC
CMS data at 7 TeV \cite{CMSJE} from the anti-$k_T$ jet algorithm
\cite{Cacciari:2008gp} in Fig.~\ref{CMSJE}, which are also
consistent with the data in various $P_T$ intervals. Since one can
separate the contributions from the light-quark jet and the gluon
jet, the comparison with the CDF and CMS data implies that
high-energy (low-energy) jets are mainly composed of the light-quark
(gluon) jets. Therefore, a precise measurement of the jet energy profile
as a function of jet transverse momentum can be used to experimentally
discriminate the production mechanism of jets in association with other
particles, such as electroweak gauge bosons, top quarks and Higgs bosons.

\section{SUMMARY}

In this article I have reviewed the resummation through the
variation of the Wilson lines off the light cone, and its applications
to the derivation of all the known single- and double-logarithm summations,
including their unifications. The idea is that the collinear dynamics
involved in a collision subprocess is independent of the Wilson line
direction $n$, so the variation effect can be factorized out of the subprocess,
leading to an evolution equation in $n$. The solution to this evolution
equation then resums important infrared logarithms in the subprocess.
For the derivations of various logarithmic summations, the point is the
treatment of the real gluon contributions to the subdiagram containing the
special vertex in the resummation formalism. Simply adopting the soft
approximations appropriate in different kinematic regions, {\it i.e.},
neglecting the $l^+$ or $l_T$ dependence in the TMD associated with the real
gluon emission, the formalism reduces to the $k_T$ resummation, the
BFKL equation, the threshold resummation, or the DGLAP equation. If keeping
both the $l^+$ and $l_T$ dependencies, the joint resummation for large $x$
and the CCFM equation for intermediate and small $x$ are obtained.
The same technique has been applied to the study of jet substructures,
and it has been shown that the invariant mass distributions and the energy
profiles of the light-quark and gluon jets can be calculated.

In this framework only the one-loop subdiagrams were evaluated for
demonstration, which corresponds to the summation of ladder graphs, or
to the summation of real gluon emissions under strong kinematic orderings.
To improve the accuracy of resummation, non-ladder graphs and
contributions from the configuration without strong kinematic orderings
need to be included. By computing the subdiagrams to two loops,
the former give next-to-leading-logarithmic corrections. The contribution
from the region without, for example, the $k_T$ ordering is taken into
account by keeping the $l_T$ dependence of the TMD, similar to the derivation
of the BFKL equation appropriate for the multi-Regge region.
That is, theoretical extensions of the resummation formalism with the Wilson
lines off the light cone are also promising.

\acknowledgments{
This work was supported by the National Science
Council of R.O.C. under Grant No. NSC-101-2112-M-001-006-MY3 and
by the National Center for Theoretical Sciences of R.O.C...}

\end{document}